\documentstyle[aps,preprint]{revtex}

\begin{document}
\draft

\title{General Static Spherical Solutions of  $d$-dimensional Charged 
         Dilaton Gravity Theories }
\author{Youngjai Kiem}
\address{Department of Physics  \\
              Sejong University \\
              Seoul 143-747, KOREA}
\author{Dahl Park}
\address{Department of Physics \\
              KAIST \\
              Taejon 305-701, KOREA}

\maketitle
\begin{abstract}
We get the general static, spherically symmetric solutions of the 
$d$-dimensional Einstein-Maxwell-Dilaton theories by dimensionally 
reducing them to a class of 2-dimensional dilaton gravity theories.  
By studying the symmetries of the actions for the static equations of
motion, we find field redefinitions that nearly reduce these theories 
to the $d$-dimensional Einstein-Maxwell-Scalar theories, and therefore 
enable us to get the exact solutions.  We do not make any
assumption about the asymptotic space-time structure.  As a result,
our 4-dimensional solutions contain the asymptotically flat 
Garfinkle-Horowitz-Strominger (GHS) solutions and the non-asymptotically
flat Chan-Horne-Mann (CHM) solutions.  Besides, we find some new solutions 
with a finite range of allowed radius of the transversal sphere.
These results generalize to an arbitrary space-time dimension 
$d$ ($d>3$).
\end{abstract}

%

\section{Introduction}

The low energy effective action of the string theory in
$d$-dimensional target space
contains the dilaton field $f$, the metric tensor
$g_{\mu \nu}^{(d)}$and a $U(1)$
gauge field, among other possible fields.  
The coupling of the dilaton field to the
$U(1)$ gauge field makes this theory different from the
Einstein-Maxwell theory.  Specifically, this coupling
is present in the action
\begin{equation}
 I = \int d^{d} x \sqrt{g^{(d)}} ( R - \frac{1}{2}
g^{(d)\alpha \beta} \partial_{\alpha} f \partial_{\beta} f
+ \frac{1}{4} e^{\chi f} F^2 )
\label{daction}
\end{equation} 
when the value of the constant $\chi$ is non-zero.
Here $F^2$ is the square of the curvature 2-form of the $U(1)$ gauge 
field.  The usual $d$-dimensional 
Einstein-Maxwell-Scalar theories correspond to (\ref{daction})
with $\chi = 0$.
The $d$-dimensional target space effective action,
derived from the condition of vanishing beta-function
from heterotic string theory, reduces to the same action with
non-vanishing $\chi$ if we set the 3-form field to be zero, after
a scale transformation of the metric.  

The action (\ref{daction}) provides us with a relatively
simple model through which we can learn the implications
of the string theory on the gravitational physics.  The dilaton
field $f$, which can either result from the conventional 
Kaluza-Klein theory or from the string theory, drastically
modifies the classical and quantum dynamics of the space-time.  
Gibbons first obtained exact black hole solutions of 
the action (\ref{daction}) in the 4-dimensional space-time
\cite{gibbons}.  Then, the further generalizations into the $d$-dimensional
case were achieved by Gibbons and Maeda (GM) \cite{gm}.
As the interest in the string theory rises, the 4-dimensional black
hole solutions were rediscovered by Garfinkle, Horowitz, and 
Strominger (GHS) in the context of the low energy effective
theory of the string theory \cite{GHS}.    
Furthermore, general static, spherically symmetric and
asymptotically flat solutions were 
obtained in Ref. \cite{rakh} for the 4-dimensional
theory and in Ref. \cite{gurses} for $d$-dimensional theory. 
These works imply the uniqueness of the GHS black holes (or 
GM black holes in $d$-dimensions) in the sense that
asymptotically flat solutions other than the GHS (or GM) black 
holes contain naked singularities.

Recently, Chan, Horne and Mann (CHM) \cite{CHM} made an interesting
point by obtaining some new 
static and spherically symmetric black hole 
solutions that do not satisfy asymptotic flatness condition,
via a special ansatz.  
Considering these developments, it is of
some interest to classify all the possible solutions of 
(\ref{daction}) without assuming the asymptotic flatness
(including the most important $d=4$ case).
In this paper, we achieve this task by obtaining
the general static spherically symmetric solutions of 
(\ref{daction}) in a local space-time region.

The assumption of the spherical symmetry is essential in our 
approach since we can then dimensionally reduce our $d$-dimensional
system to 2-dimensional one.  Thus, what we specifically solve in this
paper is the following 2-dimensional action \cite{banks} \cite{giddings}. 
\begin{equation}
I = \int d^2 x \sqrt{-g} e^{-2 \phi} [ R + \gamma g^{\alpha \beta}
\partial_{\alpha} \phi \partial_{\beta} \phi + \mu e^{2\lambda\phi} 
-\frac{1}{2} g^{\alpha \beta} \partial_{\alpha}
f\partial_{\beta} f + \frac{1}{4} e^{\epsilon \phi
+ \chi f } F^2 ],
\label{oaction}
\end{equation}
where $R$ denotes the 2-dimensional scalar curvature and 
$F$, the curvature 2-form for an
Abelian gauge field. The fields $\phi$ and $f$ represent a 
(2-dimensional) dilaton field and a massless
scalar field, respectively. The parameters $\gamma$, $\mu$, $\lambda$,
$\epsilon$ and $\chi$ are assumed to be arbitrary real numbers
satisfying the condition $2 - \lambda - \gamma /2 + \epsilon /2
= 0$\footnote{The meaning of this condition will be explained in
Sec. II.}. 
Of course, the action (\ref{oaction}) is of interest in itself
as a class of exactly solvable 2-dimensional dilaton gravity theories
coupled with a U(1) gauge field and a scalar field \cite{banks}.  
In our context, however, its importance
comes from the fact that (\ref{oaction}) is the spherically
symmetric reduction of (\ref{daction}) as 
reported, for example, in \cite{birkhoff}. 
We can relate the 2-dimensional dilaton field $\phi$ to the geometric
radius of each $(d-2)$-dimensional sphere in $d$-dimensional 
spherically symmetric space-time.  Thus, we write the
spherically symmetric $d$-dimensional metric as the sum of          
longitudinal part and transversal angular part, 
\[ ds^2 = g_{\alpha \beta} dx^{\alpha}dx^{\beta} - 
\exp {\left(- \frac{4}{d-2} \phi \right)} d \Omega, \]
where $d\Omega$ is the metric of a sphere $S^{d-2}$ with 
the unit radius and we use $(+- \cdots -)$ metric 
signature.  The spherically symmetric reduction of      
the action (\ref{daction}) 
becomes Eq. (\ref{oaction}) with $\gamma = 4 (d-3)/(d-2)$,
$\lambda = 2/(d-2)$, $\epsilon = 0$, and 
the parameter $\mu \le 0$, which depends
on the area of $(d-2)$-dimensional sphere 
after the $(d-2)$-dimensional angular integration.  Thus, 
the $d$-dimensional dilaton field $f$ becomes the scalar field 
of the 2-dimensional theory.  From the point of view of the
2-dimensional dilaton gravity theories, the
theories in our consideration continuously change from the $d=4$ 
Einstein-Maxwell-Dilaton theory to the
Callan-Giddings-Harvey-Strominger (CGHS) model as we change the
value of the real parameter $d$ from 4 to the positive 
infinity \cite{cghs}.

The method we use in section II to obtain the 
exact solutions originates from \cite{kp}.
It was applied to $d$-dimensional Einstein-Maxwell-Scalar
theories in \cite{pk}.  Just as in case of the Einstein-Maxwell-Scalar
theories where $\chi = 0$, the integration of the coupled
second order differential equations to obtain the first order
system is possible through the identification of 4 underlying
symmetries of our action (\ref{oaction}), since we have 
4 fields to solve in conformal gauge that we adopt in this
paper.  By studying these symmetries, we find a set of field
redefinitions, which nearly reduces the $\chi \ne 0$ theories
to the $\chi = 0$ theories.  These field 
redefinitions are essentially the dressing of the 2-dimensional
dilaton field and the conformal factor of the metric
with the factor $\exp ( \chi/(2a) h f)$ where 
$h$ is the scaling dimension of each field and 
$a \equiv 2 - \lambda - \gamma /4$.  They almost reduce the original
equations of motion to the ones with $\chi = 0$ with accompanying
transformations of the Noether charges.  We can directly solve
them to get the closed form expressions for our fields.
As a by-product of our method,
the relationship between the $\chi = 0$ theories and $\chi \ne 0$
theories becomes apparent, thereby telling us the influence of the
dilaton field on the space-time geometry.  

The novel point in our approach is the lack of the assumptions
about the global structure of the space-time.  As shown in
section III, our results
in 4-dimensional case contain the asymptotically
flat GHS black holes, as well as 
the CHM black holes where the
space-time is not asymptotically flat.  If the dilaton charge
changes slightly from the value needed to give a black
hole solution, naked singularities form near the would-be position
of the black hole horizon.  This behavior is consistent with
the no-hair property of a generic black hole.  It is found that the 
existence of the CHM solutions depends on the fact that $\chi \ne 0$.
If $\chi = 0$, the analog of the CHM solutions has the global
structure of ${\cal M}_2 \times S^2 $, where the transversal sphere
has a constant radius.  This structure is quite different from the
CHM solutions and the 4-dimensional spherically symmetric
Minkowski space-time.  Some new solutions in this paper, 
though we believe
they are unphysical in the end, include the space-time
geometries with the finite range of the radius of the transversal
sphere, which have other peculiar structures.  All the generic features
of the 4-dimensional results generalize to an arbitrary space-time
dimension $d$ where $d>3$.

\section{Symmetries and Derivation of Solutions}

In this section, we present a method for the derivation of the
general static solutions of the action in our consideration.
We start by reviewing the static equations of motion we should
solve. The crucial next step in our method is the construction of the
Noether charges and understanding them.  Once the symmetries
are properly understood, it is pretty straightforward to obtain the
general static solutions in a closed form.

\subsection{Static Equations of Motion}

The analysis in this paper will be performed under the choice
of a conformal gauge, where the metric tensor is given by 
\[  g_{+-}=-e^{2\rho+\gamma\phi /2}/2,~~~g_{--}=g_{++}=0. \]
We require the negative signature for a space-like coordinate and 
the positive signature for a time-like
coordinate.  The $\phi$ is included deliberately in the conformal
factor to cancel the kinetic term for the dilaton field in
Eq. (\ref{oaction}) up to total derivative terms.  The original 
action Eq. (\ref{oaction}) then
simplifies, modulo total derivative terms, to
\begin{equation}
I = \int dx^+ dx^- ( 4 \Omega \partial_+ \partial_- \rho
+ \frac{\mu}{2} e^{2 \rho} \Omega^{1-\lambda -\gamma/4}
+ \Omega  \partial_+ f \partial_- f
 - e^{\chi f-2 \rho } \Omega^{1+\gamma /4 - \epsilon /2} F_{-+}^2 ) ,
\label{conaction}
\end{equation}
where we introduce $\Omega = \exp ( -2 \phi )$ and the curvature of 
the $U(1)$ gauge field is given by
$F_{-+}=\partial_- A_+ - \partial_+ A_- $.  In obtaining the solutions,
we should supplement the equations of motion from Eq. (\ref{conaction})
with gauge constraints resulting from the choice of conformal gauge,  
\begin{equation}
\frac{\delta I}{\delta g^{\pm \pm}} = 0,
\end{equation}
where the functional derivative of $I$ in 
Eq. (\ref{oaction}) is taken.  Explicitly, they are written
as
\begin{equation}
\partial_{\pm}^2 \Omega - 2 \partial_{\pm} \rho \partial_{\pm} \Omega
+ \frac{1}{2} \Omega ( \partial_{\pm} f )^2 = 0.
\label{gconst0}
\end{equation}
The equations of motion from Eq. (\ref{conaction}) are
\begin{equation}
\partial_+\partial_-\Omega + \frac{\mu}{4} e^{2 \rho} \Omega^{1-\lambda
- \gamma/4} +\frac{1}{2}e^{\chi f-2 \rho } \Omega^{1+\gamma/4 -\epsilon/2}
F_{-+}^2=0,
\label{emrho}
\end{equation}
\begin{equation}
\partial_+\partial_-\rho + \frac{\mu}{8}(1-\lambda - \frac{\gamma}{4} )
e^{2 \rho}\Omega^{-\lambda -\gamma/4} + \frac{1}{4}
\partial_+ f \partial_- f 
\label{em0}
\end{equation}
\[ - \frac{1}{4}(1+\frac{\gamma}{4} - \frac{\epsilon}{2})
e^{\chi f-2\rho } \Omega^{\gamma/4 -\epsilon/2} F_{-+}^2 =0, \]
along with the equations for the massless scalar field,
\begin{equation}
(\partial_+\Omega\partial_-f + \partial_-\Omega\partial_+f)+2\Omega
\partial_+\partial_-f
+ \chi e^{\chi f-2\rho}\Omega^{1+\gamma/4 -\epsilon/2}
F_{-+}^2=0
\label{emf}
\end{equation}
and for the gauge fields,
\begin{equation}
\partial_-(e^{\chi f-2 \rho } \Omega^{1+\gamma/4 -\epsilon/2} F_{-+})=0,
\end{equation}
\begin{equation}
\partial_+(e^{\chi f-2 \rho } \Omega^{1+\gamma/4 -\epsilon/2} F_{-+})=0.
\label{emgf}
\end{equation}

The equations for the Abelian gauge field are immediately 
integrated to give
\begin{equation}
F_{-+}=e^{-\chi f+2 \rho } \Omega^{-1-\gamma/4 +\epsilon/2} Q ,
\label{sgauge}
\end{equation}
where $Q$ is a constant.  The absence of the emission of physical 
photons in $s$-wave Maxwell theory, due to the transversal 
polarization of physical photons, is responsible for the particularly
simple dynamics of the $U(1)$ gauge field.

The general static solutions under a particular choice
of the conformal gauge can be found by assuming that the metric $\rho$,
the scalar field $f$, and $\Omega$ depend only on a single space-like 
coordinate $x=x^+ + x^-$.  Then, Eq. (\ref{sgauge}) implies the 
curvature 2-form $F_{-+}$ also depends only on $x$.  Therefore,
from here on,
we introduce a field $A(x)$ that depends only on $x$ and satisfies
\[ F_{-+} = \frac{dA}{dx} .  \]
In fact, the choice of the vector potential $A_{\pm} = \pm A /2$,
gives the same expression for the curvature 2-form.
The original equations of motion, Eqs. (\ref{emrho})--(\ref{emgf}),
then consistently reduce to a system of the following
coupled second order ordinary
differential equations (ODE's).
\begin{equation}
\Omega^{\prime \prime} + \frac{\mu}{4} e^{2 \rho} \Omega^{1-\lambda
-\gamma/4} +\frac{1}{2}e^{\chi f-2 \rho } \Omega^{1+\gamma/4 -\epsilon/2
} A^{\prime 2} =0,
\label{eomrho}
\end{equation}
\begin{equation}
\rho^{\prime \prime} + \frac{\mu}{8}(1-\lambda -\frac{\gamma}{4})
e^{2 \rho} \Omega^{-\lambda -\gamma/4} + \frac{1}{4}
f^{\prime 2 } - \frac{1}{4}(1+\frac{\gamma}{4} -\frac{\epsilon}{2}) 
e^{\chi f-2 \rho }
\Omega^{\gamma/4 -\epsilon/2} A^{\prime 2} =0,
\end{equation}
\begin{equation}
\Omega f^{\prime \prime} +  \Omega^{\prime} f^{\prime} + \frac{\chi}{2}
e^{\chi f-2\rho}\Omega^{1+\gamma/4 -\epsilon/2} 
A^{\prime 2} =0,
\end{equation}
and
\begin{equation}
\frac{d}{dx} \left( e^{\chi f-2\rho} \Omega^{1+\gamma/4 -\epsilon/2}
A^{\prime}  \right)=0,
\label{eoma}
\end{equation}
where the prime represents taking a derivative with respect to $x$.
The general solutions of the above ODE's are the same as the general static
solutions of the original action under a particular choice of the conformal
coordinates. They can be derived from the action
\begin{equation}
I = \int dx [ \Omega^{\prime} \rho^{\prime} - \frac{\mu}{8}
e^{2 \rho}\Omega^{1-\lambda -\gamma/4}
-\frac{1}{4} \Omega  f^{\prime 2} 
+\frac{1}{4} e^{\chi f-2 \rho} \Omega^{1+\gamma/4 -\epsilon/2} 
A^{\prime 2} ] .
\label{xaction}
\end{equation}
The gauge constraints reduce to 
\begin{equation}
\Omega^{\prime \prime} -2 \rho^{\prime} \Omega^{\prime}
+\frac{1}{2}\Omega f^{\prime 2} =0,
\label{gconst}
\end{equation}
if we restrict our attention to static solutions.  
Eqs. (\ref{xaction}) and (\ref{gconst}) are the starting point
for our further considerations on general static solutions.

\subsection{Symmetries, Noether Charges and Field Redefinitions}

We observe the following four continuous symmetries of the action
Eq. (\ref{xaction})
\begin{eqnarray*}
(a) \ \ &&f \rightarrow f + \alpha, \ A \rightarrow A e^{-\chi\alpha/2},\\
(b) \ \ &&A \rightarrow A + \alpha,\\
(c) \ \ &&x \rightarrow x + \alpha,\\
(d) \ \ &&x \rightarrow e^{\alpha}  x, \ 
     \Omega \rightarrow e^{\alpha} \Omega, \
     e^{2 \rho} \rightarrow e^{ - (2-\lambda - \gamma/4) \alpha} 
     e^{2 \rho}, \
     A \rightarrow e^{ - (1 - \lambda /2 - \epsilon /4 ) \alpha} A, 
\end{eqnarray*}
where $\alpha$ is an arbitrary real parameter of each transformation.
The Noether charges for these symmetries are constructed as
\begin{eqnarray}
f_0 & = &\Omega f^{\prime}  
+ \frac{\chi}{2}e^{\chi f-2\rho}
\Omega^{1+\gamma /4 - \epsilon /2} A  A^{\prime},
\label{f0} \\
Q  & =&  e^{\chi f-2 \rho} \Omega^{1+\gamma /4 -\epsilon /2} 
A^{\prime}, \label{a0} \\
c_0 & = &\Omega^{\prime} \rho^{\prime} - \frac{1}{4}\Omega  f^{\prime 2} 
+ \frac{1}{4}e^{\chi f-2\rho} \Omega^{1+\gamma /4 - \epsilon /2} A^{\prime 2} 
+ \frac{\mu}{8} e^{2 \rho} \Omega^{1 - \lambda - \gamma /4}, \label{c0} \\
s + c_0  x & = &-\frac{1}{2} (2-\lambda-\frac{\gamma}{4}) \Omega^{\prime}
+ \rho^{\prime} \Omega 
- \frac{1}{2} (1- \frac{\lambda}{2} - \frac{\epsilon}{4} ) e^{\chi f-2\rho}
\Omega^{1+\gamma /4 - \epsilon /2} A  A^{\prime}, \label{c1}
\end{eqnarray}
respectively.  The gauge constraint, Eq. (\ref{gconst}), implies
$c_0 = 0$.

We compare these symmetries to the symmetries in
Ref. \cite{pk} where one considers the case of 
$\chi = 0$\footnote{In \cite{pk}, we chose different conformal
coordinates.  Our choice in this paper is related to the one in
\cite{pk} via a conformal transform $x^{\pm} \rightarrow \ln
x^{\pm}$.}.
The symmetries $(b)$, $(c)$ and $(d)$ are exactly the same in
both cases.  The symmetry $(b)$ results from the trivial dynamics
of the $U(1)$ gauge field in $s$-wave theories.  The symmetries $(c)$
and $(d)$ are remnants of the underlying classical conformal symmetries  
of the original action (\ref{oaction}).  The only difference 
between our case and Ref. \cite{pk} lies in 
the symmetry $(a)$.  Due to the coupling
of $f$ to the gauge field (the $\exp (\chi f)$ factor in Eq. 
(\ref{xaction})), the translation of the $f$ field should be compensated
by an additional scale transformation of $A$, to make the action
Eq. (\ref{xaction}) invariant.    

The symmetry $(d)$ defines the scaling dimension of each field.  From
here on, we will impose a condition
$ a \equiv 2-\lambda - \gamma / 4 = 1- \lambda /2 - \epsilon  /4 
\ne 0 $, 
which enables us
to straightforwardly solve the equations of motion to get the
solutions in a closed form.  This condition means the scaling 
dimension of the $A$ field and that of the $\exp (2 \rho )$ are the same.
We note all $s$-wave reduction of the $d$-dimensional 
Einstein-Maxwell-Dilaton theories satisfies this condition and 
gives the value $a = (d-3)/ (d-2) $. 

The difference between the $\chi = 0$ case and the non-zero $\chi$ case
is the additional scale transformation of the $A$ field in the
symmetry $(a)$.  However, we can adjoin $(d)$ to $(a)$ to make $A$
invariant.  Thus, we consider such an adjoined symmetry
\begin{equation}
(a)^{\prime} \ \  x \rightarrow e^{- \chi \alpha / (2a) }  x, \ 
     \Omega \rightarrow e^{- \chi \alpha / (2a) } \Omega, \
     e^{2 \rho} \rightarrow e^{ \chi \alpha /2 } 
     e^{2 \rho}, \
     A \rightarrow A, \ f \rightarrow f + \alpha ,
\end{equation}
where we made a certain $(d)$ transformation following the $(a)$
transformation to make $A$ invariant.  Now as far as $f$ and $A$
fields are concerned, the transformations in $(a)^{\prime}$ are 
independent of
the value of $\chi$.  For other fields, we introduce the following
field redefinition 
\begin{eqnarray}
\bar{\Omega} & = & \exp ( \frac{\chi}{2a} f) \Omega ,   \nonumber \\
e^{2 \bar{\rho }} & = & \exp ( - \frac{\chi}{2} f )  e^{2 \rho} ,
\nonumber \\
 d \bar{x} & = & \exp ( \frac{\chi}{2a} f ) dx .
\label{fredef}
\end{eqnarray}
Then, written in terms of these redefined fields, we have
\begin{equation}
(a)^{\prime} \ \  \bar{x} \rightarrow  \bar{x}, \ 
     \bar{\Omega} \rightarrow \bar{\Omega}, \
     e^{2 \bar{\rho}} \rightarrow e^{2 \bar{\rho}}, \
     A \rightarrow A, \ f \rightarrow f + \alpha
\end{equation}
for the transformation $(a)^{\prime}$.   The symmetry $(b)$ obviously
survives the field redefinition.  Since the original action
and the field redefinition does not have explicit dependence on $x$, 
the symmetry (c) remains intact.  The symmetry $(d)$
still holds even if we change $(\rho , \Omega )$
into $(\bar{\rho} , \bar{\Omega} )$, since the $f$ field does not
change under the symmetry $(d)$.  Thus, in terms of the redefined fields,
the symmetries $(a)^{\prime}$, $(b)$, $(c)$ and $(d)$ do not 
depend on the value of $\chi$, and look identical to the symmetries of
Ref. \cite{pk} where one considers the $\chi = 0$ case.  
Thus, a natural expectation is
that if we use the redefined fields to solve the equations of motion,
the analysis will be similar to the $\chi = 0$ case.

\subsection{Explicit Solutions}

In terms of redefined fields, we rewrite Eqs. (\ref{f0}) and
(\ref{a0}) as
\begin{equation}
f_0 = \bar{\Omega} \dot{f} + \frac{1}{2}\chi QA ,
\label{tf0}
\end{equation}
\begin{equation}
Q = e^{ -2 \bar{\rho}} \bar{\Omega}^{a+1} \dot{A} ,
\label{ta0}
\end{equation}
where the overdot represents the differentiation with respect to
$\bar{x}$.  We notice that, unlike Eq. (\ref{a0}), the $\chi$
dependence is absent in Eq. (\ref{ta0}).   After the field
redefinition and using Eq. (\ref{tf0}), Eq. (\ref{c1}) becomes
\begin{equation}
\bar{s} = - \frac{a}{2} \dot{\bar{\Omega}} + 
\dot{\bar{\rho}} \bar{\Omega} - \frac{a}{2} \bar{Q} A .
\label{tc1}
\end{equation}
All the $\chi$ dependence can be absorbed into the redefinition of the 
parameters given by
\[ \bar{s} = s - \frac{\chi}{2} f_0  \ , \  
   \bar{Q} = \frac{ a + \chi^2 /2 }{a} Q . \]
If we set $\chi = 0$, there is no change other than 
the disappearance of the overbars in Eq. (\ref{tc1}).
In our further consideration, we only consider the case 
when $Q \ne 0$\footnote{If the U(1) charge vanishes, the whole 
situation becomes identical to that
of Ref.\cite{kp}, where we already have a complete analysis.}.
By this assumption, we exclude the case 
when the $A$ field becomes degenerate,
being a strict constant.  Then, we combine Eqs. (\ref{ta0}) and 
(\ref{tc1}) to get
\[ \frac{d}{d\bar{x}} ( Q e^{2\bar{\rho}} \bar{\Omega}^{-a} )  =
 ( 2 \bar{s} +  a \bar{Q} A ) \dot{A} , \]
which can be integrated to yield
\begin{equation}
 Q e^{2\bar{\rho}} \bar{\Omega}^{-a} =
 2 \bar{s} A + \frac{a}{2} \bar{Q} A^2+ c \equiv P(A) ,
\label{Aor}
\end{equation}
where $c$ is the constant of integration.
After the field redefinition and upon using Eqs. (\ref{tf0}),
(\ref{ta0}) and (\ref{Aor}), we rewrite Eq. (\ref{c0}) as
\begin{equation}
0 = \dot{\bar{\Omega}} \dot{\bar{\rho}} 
+ \frac{\mu}{8} e^{ 2 \bar{\rho}} \bar{\Omega}^{a-1} 
- \frac{1}{4} \frac{\bar{f}_0^2}{\bar{\Omega}}
+ \frac{1}{4} \bar{Q} \dot{A} ,
\label{tc0}
\end{equation}
where we introduced
\[ \bar{f}_0^2 = \frac{a - \chi^2 /2 }{a} f_0^2 
+ \frac{2 \chi }{a} sf_0 +  \frac{\chi^2}{2a} Q c . \]
For $\chi = 0$, we get $\bar{f}_0^2 = f_0^2$.  As before,
the disappearance of the overbars is the only 
change in Eq. (\ref{tc0}) as 
we set $\chi = 0$.  As expected from the previous discussions,
the expressions for the Noether charges, Eqs. (\ref{ta0}),
(\ref{tc1}), and (\ref{tc0}), become independent of $\chi$,
modulo the redefinitions of the parameters and fields. 

Although the Noether charge
expression for the symmetry $(a)$, Eq. (\ref{tf0}), still contains
$\chi$ dependence, it is simple enough for us to give an exact analytic
treatment.  Specifically, from Eqs. (\ref{tf0}), (\ref{ta0}) and 
(\ref{Aor}), we can determine $f$ via
\begin{equation}
\dot{f}=\frac{f_0-\chi Q A /2}{P(A)}\dot{A}
\label{fdot}
\end{equation}
which, upon integration, becomes
\begin{equation}
f(A)=\frac{a f_0 +  \chi s }{a + \chi^2 /2 }I(A)-
\frac{\chi /2}{a + \chi^2 /2 } \ln{|P(A)|}+ f_1 ,
\label{fs}
\end{equation}
where $f_1$ is the constant of integration and we introduce
$I(A) = \int P(A)^{-1} dA$.

Using Eqs. ({\ref{ta0}) and (\ref{Aor}), we can rewrite
Eq. (\ref{tc0}) as
\begin{equation}
0=4a \left(\frac{d\bar{\phi}}{dA}\right)^2
-\frac{2}{P(A)}\frac{dP(A)}{dA} \frac{d\bar{\phi}}{dA}
-\frac{\bar{f}_0^2} {2P^2(A)}
+\frac{2 \bar{Q} +\mu e^{-4a \bar{\phi}}/\bar{Q}}{4P(A)} ,
\label{OaA}
\end{equation}
where we introduced $\bar{\phi}$ via $\bar{\Omega} = 
(a/(a+ \chi^2 /2 ))^{1/(2a)} 
\exp (-2 \bar{\phi} )$.   
By differentiating the above
equation with respect to $A$, we have
\begin{equation}
0=\left[ \frac{1}{P}\frac{dP(A)}{dA} - 4a\frac{d\bar{\phi}}{dA} \right]
\left[ \frac{d^2\bar{\phi}}{dA^2} +2a \left(\frac{d\bar{\phi}}{dA} \right)^2
-\frac{\bar{f}_0^2}{4P^2} \right].
\label{twofac}
\end{equation}
Eqs. (\ref{OaA})) and (\ref{twofac}) are exactly 
the same as Eqs. (31) and (32) of
Ref. \cite{pk} where one considers the $\chi =0$ case, except 
the overbars on the
variables. Thus, we can follow that paper to obtain 
the exact solutions.
The form of the solutions depends on a parameter $D$ defined by
\[  D^2 \equiv 4 \bar{s}^2 - 2a  \bar{Q} c + 2a \bar{f}_0^2 
            = 4 s^2 -2a  Qc + 2a f_0^2  .\]
Here one interesting observation is about the second identity.  
We can easily
check that the $\chi$ dependence in $D$ drops out as 
the identity shows.

When the second factor in Eq. (\ref{twofac}) vanishes, we have
\begin{equation}
e^{-4a \bar{\phi} }=\frac{8D^2 \bar{Q} }{-\mu}\frac{c_1 e^{D I}}
{P \left[c_1 e^{D I}-2 a \right]^2}   
\label{s1}
\end{equation}
for $D^2 > 0 $,
\begin{equation}
e^{-4a \bar{\phi}}= \frac{4 \bar{Q}}{- \mu a } \frac{1}{P(I+c_1)^2}
\label{s2}
\end{equation}
for $D^2 = 0$ and
\begin{equation}
e^{-4a \bar{\phi} } =  \frac{ \tilde{D}^2 \bar{Q}}{- \mu a}
\frac{1}{\sin^2 ( \tilde{D} I /2 + c_1 ) }
\label{s3}
\end{equation}
for $D^2 = -\tilde{D}^2 <  0$,
where $c_1$ is the constant of integration.  If $P(A) = 0 $ has 
two real roots
$A = \bar{A}_-$ and $A = - \bar{A}_+$, we have
\[ e^{DI} = \Bigg| \frac{ A- \bar{A}_-}{A + \bar{A}_+} 
                \Bigg|^{D/ \sqrt{ 4 \bar{s}^2 - 4a \bar{Q} c}} \]
and 
\[ P(A) = \frac{a}{2} \bar{Q} ( A + \bar{A}_+ ) ( A -  \bar{A}_- ) . \]
There are two other expressions for $\exp ( DI)$
depending on the sign of $4 \bar{s}^2 -2a \bar{Q} c$; when 
$P(A)$ has a double root
and when $P(A)$ has two complex roots, respectively.

When the first factor in 
Eq. (\ref{twofac}) vanishes, we have 
\begin{equation}
e^{-4a \bar{\phi}} =\frac{ D^2 \bar{Q}}{\mu a}\frac{1}{P}.
\label{ops}
\end{equation}
This solutions does not contain a further constant of integration.

Since $Q  \ne 0 $, we find $A$ as a function of $x$ by plugging
Eq. (\ref{Aor}) into Eq. (\ref{ta0}),
\begin{equation}
 \bar{x} - \bar{x}_0 = \int \frac{\bar{\Omega}(A)}{P(A)}dA,
\label{Ax}
\end{equation}
where $\bar{x}_0$ is the constant of integration.  
Eqs. (\ref{Aor}), (\ref{fs}), (\ref{s1}) - (\ref{ops}) and
(\ref{Ax}) represent explicit expressions for the general solutions 
in terms of the redefined fields.  

\section{Aspects of General Static Solutions}

Our derivation of the general static solutions does not sensitively
depend on the value of the parameter $a$, as long as the restriction
$a>0$ holds.  Thus, we will give a unified description of the properties 
of the general static solutions for an arbitrary space-time 
dimension $d$, which is larger than 3.  When interpreted as the 
dimensionally reduced 2-d dilaton gravity theories, the parameter $d$
can even be considered as a continuous parameter.  
Being the general solutions, our results are somewhat complicated
and the solution space has a quite rich structure.  As a result, we
will first concentrate on a class of solutions near to the GHS solutions
and CHM solutions in $d$-dimensional space-time with an arbitrary
dilaton coupling constant $\chi$.   Then, we will discuss the solutions
presented in Eq. (\ref{ops}).

\subsection{GHS and CHM Solutions in $d$ dimensions}

The assumption of the asymptotic flatness plays an important role
in many literature concerning the derivation of the static solutions
in $d$-dimensional Einstein-Maxwell-Dilaton theories with an
arbitrary or a fixed value of 
$\chi$ \cite{gibbons}\cite{gm}\cite{GHS}\cite{rakh}\cite{gurses}.  
These kinds of investigation lead to the GHS solutions 
(or Gibbons-Maeda solutions)
in $d$-dimensional space-time.  However, the discovery of CHM
black holes \cite{CHM} shows there are some interesting non-asymptotically
flat solutions.  One of the main purposes of this subsection is
to illuminate the relationship between the CHM and the GHS solutions,
since our solutions include both of them and show how they are continuously
connected.

Throughout the presentation in this subsection,
we attempt to show the relationship 
between the $\chi = 0$ solutions and the $\chi \ne 0$ solutions.  In 
case of $\chi = 0$ theories, the no-hair theorem prevents the non-zero value 
of the scalar charge.  It is possible to verify this theorem by showing 
that any non-zero value of the scalar charge produces a naked singularity.  
We find there is essentially the same phenomenon in case of $\chi \ne 0$.
   
Among the general static solutions we obtained in Sec. II, the solutions
described by Eq. (\ref{s1}) depend on 7 independent 
parameters, $f_0$, $f_1$, $Q$, $c$, $s$, $c_1$, and $\bar{x}_0$, 
satisfying $D^2 >0$. We note each parameter with overbar reduces to the 
one without overbar if we set $\chi = 0$.
We expect 8 constants of motion from solving the second 
order differential equations for 4 fields.  
One of them is set to zero by the gauge 
constraint for the choice of a conformal gauge.  
In order to make contact with
aforementioned theories, we further impose two conditions, $Q \ne 0$ and
$4\bar{s}^2-2a\bar{Q}c>0$, for the constants of motion. Under these
conditions, from Eq. (\ref{Aor}) we have
\begin{equation}
2e^{2\rho}\Omega^{-a}=(a+\chi^2/2)e^{\chi f}(A+\bar{A}_+)
(A-\bar{A}_-)
\label{metric}
\end{equation}
where we introduced
\[ \bar{A}_{\pm}=(\sqrt{4\bar{s}^2-2a\bar{Q}c}\pm2\bar{s})/(a\bar{Q}) .\]
Here, $e^{2\rho}\Omega^{-a}$ corresponds to the conformal factor of the
longitudinal part of the metric. For the dilaton field, Eq. (\ref{fs})
yields
\begin{equation}
f(A)=f_+ \ln |A+\bar{A}_+| -f_- \ln |A-\bar{A}_-|
-\frac{\chi/2}{a+\chi^2/2}\ln|a\bar{Q} /2|+f_1 ,
\label{dilaton}
\end{equation}
where we defined two parameters $f_{\pm}$ via
\[ f_{\pm}=(f_0\pm \chi Q \bar{A}_{\pm} /2)/\sqrt{4\bar{s}^2-2a\bar{Q}c}.\]
The $\Omega$ field can be obtained from Eq. (\ref{s1}) as follows.
\begin{equation}
\Omega^{2a}= C \frac{|A+\bar{A}_+|^{h_+}|A-\bar{A}_-|^{h_-}}
{(c_1|A-\epsilon\bar{A}_{-\epsilon}|^{\alpha}
-2a|A+\epsilon\bar{A}_{\epsilon}|^{\alpha})^2}
\label{omega}
\end{equation}
where $\alpha=|D|/\sqrt{4\bar{s}^2-2a\bar{Q}c}>0$, $\epsilon=D/|D|$,
$h_{\pm}=\alpha-1\pm\chi f_{\pm}$ and
\[ C =\frac{8D^2 c_1 |Q| }{-\mu}\frac{e^{-\chi f_1}}
{ | a\bar{Q} /2 |^{a/(a+\chi^2/2)}}.\]
We note once again that $\mu < 0$ for the spherically symmetric 
$d$-dimensional Einstein-Maxwell-Dilaton theories when $d>3$.
The relation between the $A$ field and the coordinate $x$ is given
by Eq.(\ref{Ax}) as
\begin{equation}
dx = \frac{\Omega}{P(A)} dA  .
\label{xcoor}
\end{equation}
For our further consideration, we set $\epsilon = +$ and $c_1 > 0$.  
We also consider
the case when $Q > 0$ and $ 0 < k \le 1$ where 
$k = (2a/ c_1 )^{1/\alpha}$.  This gives an 
inequality $\bar{A}_- > - \bar{A_+}$.  From Eq. (\ref{metric}), the physical
range of $A$ is restricted to satisfy $A \ge \bar{A}_-$ or 
$A \le - \bar{A}_+$
in order to make the left hand side positive semidefinite.

Our solutions are defined on a local space-time region.  The natural 
places to regard as boundaries of a space-time region are $\Omega = 0$,
where the transversal sphere collapse to a point, and 
$\Omega = \infty$, which corresponds to the spatial infinity.  Thus,
we start from finding the zeros and infinities of $\Omega$ given 
in Eq.(\ref{omega}).  If $\chi = 0$, we immediately see 
$h_- = h_+ = 0$ if and only if $f_0 = 0$,  and  
both of $h_{\pm}$ are positive-definite
for other values of $f_0$.  The similar story
holds in case of $\chi \ne 0$.  We have an identity
\[ h_+ h_- = \frac{a}{a + \chi^2 /2} 
   ( \frac{|D|}{\sqrt{4\bar{s}^2 - 2a \bar{Q} c }} -1 )^2 \ge 0 . \]
We observe $h_- h_+ = 0 $ if and only if $\bar{f}_0 = 0$.  If
$\bar{f}_0 \ne 0$, the product should be larger than zero.  
Furthermore, if $\bar{f}_0 = 0$, we have 
$\{ h_+ , h_- \} = \{ 0 , \chi^2 / ( a + \chi^2 /2 )\} $ and, as a result,
one of $h_{\pm}$ vanishes and the other becomes larger than zero
for a non-zero value of $\chi$.  Since $h_{\pm}$ are continuous functions
of $\bar{f}_0$, we infer that $h_{\pm}$ are positive definite 
if $\bar{f}_0 \ne 0$.  Given this information it is easy to see
the zeros and infinities of $\Omega$ given in Eq. (\ref{omega}).

For $0 < k < 1$, $\Omega$ becomes infinity when the denominator 
of Eq. (\ref{omega}) vanishes.  Specifically, there are two
values of $A$, 
$A = \bar{A}^I_{\infty} = ( \bar{A}_- - k ( - \bar{A}_+ ) )/(1-k)$
and 
$A = \bar{A}^{II}_{\infty} = ( \bar{A}_- + k ( - \bar{A}_+ ) )/(1+k)$,
for which the condition is met.
The value $\bar{A}^{II}_{\infty}$ lies between $-\bar{A}_+$ and
$\bar{A}_-$, and satisfies $(A_- - \bar{A}^{II}_{\infty} )
= k ( \bar{A}^{II}_{\infty} - ( -\bar{A}_+) )$.
The value $\bar{A}^{I}_{\infty}$ is larger than the both
of $\bar{A}_-$ and $- \bar{A}_+$, and
satisfies $(\bar{A}^{I}_{\infty} - \bar{A}_-  ) 
 = k ( \bar{A}^{I}_{\infty} - ( - \bar{A}_+ ) ) $.  

The structure of the
zeros depends on whether $\bar{f}_0 = 0$ or not.  For $\bar{f}_0 = 0$
case, we consider $h_- = 0$ and $h_+ = \chi^2 / (a + \chi^2 /2 )$,
thereby resulting $\alpha = 1$.
Then there are three zeros of $\Omega$ at $A = - \infty$,
$A = - \bar{A}_+$ and $A = \infty$.  It turns out that the GHS 
(or GM) black holes outside the outer horizon belong to this case.  
The range of $A$ that should make a 
invertible map between $A$ and $\Omega$ is chosen to be 
$[ \bar{A}_- , \bar{A}^{I}_{\infty} )$.  The limit $A = \bar{A}_-$
is special since for this value of $A$ and under the above
conditions, we find
\[
\partial_+ \Omega = \frac{d \Omega}{dx} 
= \frac{\bar{Q}}{4} [ h_- (A+ \bar{A}_+ ) + h_+ ( A - \bar{A}_- )
\]
\begin{equation}
- 2 \alpha 
\frac{ |A+ \bar{A}_+ |^{\alpha} (A-\bar{A}_- ) 
      - k^{\alpha} |A - \bar{A}_- |^{\alpha} (A + \bar{A}_+ ) }
     { |A+ \bar{A}_+ |^{\alpha} 
      - k^{\alpha} |A - \bar{A}_- |^{\alpha}  } ]
\label{domega}
\end{equation}
becomes zero, signaling the existence of the apparent horizon,
which becomes the event horizon when considering static solutions.
Indeed, for $\bar{A}_- < A < \bar{A}^{I}_{\infty}$, 
$\partial_+ \Omega$ monotonically increases.  In terms of the 
geometric gauge, the solutions can be rewritten as
\[
ds^2 =  (1- \frac{r_+}{r} ) ( 1- \frac{r_-}{r} 
)^{\frac{a-\chi^2 /2}{a + \chi^2 /2 }}   dt^2 
\]
\begin{equation}
- \frac{1}{- \mu a}  ( 1- \frac{r_+}{r} )^{-1} 
( 1- \frac{r_-}{r} ) ^{- \frac{a- \chi^2 ( 1/a - 3/2 )}
                                         {a + \chi^2 /2 }}
r^{ \frac{2}{a} - 4} dr^2
- r^{\frac{1-a}{a}} (1- \frac{r_-}{r} 
)^{\frac{1-a}{a} \frac{\chi^2}{a + \chi^2 /2} } d \Omega
\end{equation}
for the metric and
\begin{equation}
e^{\chi f} = e^{\chi f_{\infty} } ( 1 - \frac{r_-}{r} )^{- \chi^2 
/ ( a + \chi^2 /2 ) }
\end{equation}
for the dilaton field.  Here the parameters $r_+$, $r_-$ and
$f_{\infty}$, the value of the dilaton field at the spatial infinity, are 
related to the black hole
mass $M$, the electric charge $Q_E$ and the dilaton charge $Q_D$
as follows.
\begin{equation}
 2 M = \frac{a - \chi^2 /2 }{a  + \chi^2 / 2 } r_- + r_+  \ , \
 Q_E = ( \frac{r_+ r_- e^{- \chi f_{\infty} }  }        
                        { 2a + \chi^2 } )^{1/2}           \ , \
 Q_D = \frac{\chi^2}{a + \chi^2 /2 } r_-  
\end{equation}
These relations are derived from the asymptotic behavior of
each field.
In terms of our constants of motion, we have the expressions
\[ r_+ = 2a \sqrt{\frac{ e^{-\chi f_{\infty} } }
{- \mu c_1 ( a+ \chi^2 /2)} }  \bar{Q} \  ,   \
r_- =  \sqrt{ \frac{c_1 e^{- \chi f_{\infty} } }
{ - \mu ( a + \chi^2 /2 ) } } \bar{Q} \]
and
\[ e^{\chi f_{\infty} } = e^{\chi f_1} 
( \frac{c_1}{c_1 - 2a} \frac{a \bar{Q} /2 }
{\sqrt{4 \bar{s}^2 - 2 a \bar{Q} c}} )^{- \chi^2 / ( a + \chi^2 / 2 ) } .\]
These results reproduce the GHS black hole
\cite{GHS} for $d=4$, i.e., $a = 1/2$ and $\mu = -2$, 
and, for $d>4$, become
the GM black holes \cite{gm}.  Physically distinctive
solutions are parameterized by 4 parameters that represent
the black hole mass, the electric charge, the dilaton charge,
and the value of the dilaton at the spatial infinity. 
Moreover, there is a 
relation that gives the dilaton charge in terms of the electric
charge, the black hole mass and the value of the dilaton field at
the spatial infinity, resulting from $\bar{f}_0 = 0$.  This 
reproduces Eq. (10) of \cite{GHS}, which plays the equivalent 
role.  The   
original parameter space is 7 dimensional, but the degrees of
freedom in choice of the reference time, the reference scale,
and in the addition of an arbitrary real number to 
the $A$ field (we recall
$dA/dx$ is related to the electric field) produce 3 dimensional
orbit.  The quotient space of the original parameter space by
this orbit results the 4 dimensional parameter space.
The condition $\bar{f}_0 = 0$ selects the 3 dimensional subspace
of independent parameters immersed in the 4 dimensional space.
It is interesting to note that the no-hair theorem in case of 
$\chi = 0$ prohibits the black hole from carrying the scalar charge,
i.e., $f_0 = 0$ for the black hole solutions, while we have 
$\bar{f}_0 = 0$ for  $\chi \ne 0$.

For $\bar{f}_0 \ne 0$, we have 4 zeros of $\Omega$ at
$A = - \infty$, $A = -\bar{A}_+$, $A = \bar{A}_-$ and
$A = \infty$ since $h_{\pm} > 0$.  
The positions of the infinities of $\Omega$ are
the same as the $\bar{f}_0 = 0$ case.  We consider what
happens to the solutions for the same range of 
$A$ that reproduces the GHS (or GM) solutions, 
$[\bar{A}_- , \bar{A}^{I}_{\infty} )$.  In this case, 
$A = \bar{A}_-$, the would-be position for the black hole horizon
just becomes the point $\Omega = 0$.  Furthermore, we have 
$\partial_+ \Omega = h_- \sqrt{4\bar{s}^2 - 2a \bar{Q} c}  / (2a) > 0$
for $A = \bar{A}_-$ from Eq. (\ref{domega}) and, as $A$ increases up to 
$A = \bar{A}^{I}_{\infty}$, the function $\partial_+ \Omega$
monotonically increases.  Thus, at $A = \bar{A}_-$, we have 
a naked singularity in contrary to the case when $\bar{f}_0 = 0$.
This behavior becomes more understandable if we compute
\[ f_+ f_- = \frac{a}{a + \chi^2 /2} 
             \frac{\bar{f}_0^2}{(4 \bar{s}^2 -2 a \bar{Q}c )} . \]
Thus, if $\bar{f}_0 \ne 0$, both of $f_{\pm}$ becomes non-zero.
This implies the dilaton field $f$ diverges logarithmically
at $A = \bar{A}_-$, as can be explicitly seen from Eq. (\ref{dilaton}).
In other words, if the dilaton field $f$ carries the dilaton charge
other than the one that satisfies $\bar{f}_0 = 0$, the dilaton field
diverges logarithmically near the position of the would-be horizon.
The stress-energy induced by this divergence cuts the space-time
off at that position and makes the would-be horizon the point of
a naked singularity.  If $\chi = 0$, we go back to the usual story
of the No-scalar-hair theorem for the charged black holes. \cite{pk} 
If $\chi \ne 0$, we have a similar story but $\bar{f}_0 = 0$ gives
non-vanishing value of $f_0$.

Our next concern is the case when $k =1 $.  As before we first 
study the case when $\bar{f}_0 = 0$ and choose to consider 
$h_- = 0$ and $h_+ = \chi^2 / ( a + \chi^2 /2 )$.  The constant
$\alpha$ is again 1 for $\bar{f}_0 = 0$.  The most distinctive
change now from the case $0 < k <1$ is the location of 
the infinities of the $\Omega$ field.  The 
$\bar{A}^{II}_{\infty}$ is now the average value of the 
$\bar{A}_-$ and $- \bar{A}_+$.  However, $\bar{A}^{I}_{\infty}$
gets pushed to the infinity.  Thus, $\Omega$ becomes infinity at
$A = - \infty$, $A = (\bar{A}_- - \bar{A}_+ ) /2$
and $A = \infty$.  The zeros of $\Omega$ is located at
$A = - \bar{A}_+$.  From this consideration, it is clear that
the range we should choose to get the analog of the GHS black
holes is $[\bar{A}_- , + \infty )$.  Eq. (\ref{domega})
once again shows $\partial_+ \Omega$ is zero at $A = \bar{A}_-$
and increases monotonically from there as $A$ increases.
Thus, $A = \bar{A}_-$ can play the role of the black hole horizon.
By introducing the geometric gauge, our solutions 
in this region
can be written as
\begin{equation}
ds^2=U(r)dt^2-U^{-1}(r)dr^2-r^{2N}d\Omega
\end{equation}
for the metric and
\begin{equation}
e^{\chi f}=\frac{2Q^2}{(d-2)(d-3)}\frac{a+\chi^2/2}{a}
r^{-2N(d-3)}  
\end{equation}
for the dilaton field.  Here we use the formula $\mu=-(d-2)(d-3)$
and assume $\chi \ne 0$.  The function $U(r)$ is given by
\[
U(r)=\frac{\chi^4}{4N^2(a+\chi^2/2)^2} r^{2(1-N)}
-\frac{4M}{(d-2)N} r^{(1-\chi^2/(2a))(1-N)},
\]
and the number $N$ satisfies $N=\chi^2  / (2(d-3)a+\chi^2 )$.
In terms of our constants of motion, the parameter $M$ is
given by
\[ 
M=\frac{\chi(s\chi+af_0)}{2a(a+\chi^2/2)}. \]
We reproduce the 
CHM black holes \cite{CHM} by transforming $f$ and $\chi$ via
$f \rightarrow f \sqrt{8/(d-2)}$, $\chi \rightarrow \chi \sqrt{2/(d-2)}$.

The counting of the physical degrees of freedom is interesting
in this case.  The degrees of freedom in choice of the reference
scale \footnote{ In \cite{CHM}, the degree of freedom in the choice
of the reference scale is represented by the parameter $\gamma$.
In their Eq. (3.2), we can redefine $\gamma^2 M \rightarrow M$
and $t / \gamma^2 \rightarrow t$.  Then, the metric and the
dilaton field depend only on two parameters $Q$ and $M$.},
the reference time, and in the translation of $A$ reduce the
physical degrees of freedom to 4.  The further imposition of
2 conditions, $\bar{f}_0 = 0$ and $k = 1$, leaves us the
2-dimensional space for the physically independent parameters.
In fact, the CHM solutions have two
independent parameters $Q_E$ and $M$,
and the dilaton field $f$ contains no further 
independent parameters.  In case of GHS solutions, we are able
to freely set the value of the dilaton field at the spatial infinity.

What happens to the CHM solutions when $\bar{f}_0 \ne 0$ is similar
to the case of generic black holes.  For $\bar{f}_0 \ne 0$,
we find that $\Omega = 0$ at $A = \bar{A}_-$ and 
$\partial_+ \Omega $ is positive definite for $A \ge \bar{A}_-$,
since $h_{\pm} > 0$.
Thus, if the CHM black holes try to carry other values of the 
dilaton charge than the one required by $\bar{f}_0 = 0$, naked
singularities occur.  Thus, we see that the story similar to the
no-hair theorem is also working here.  

Another interesting issue
is what happens to the CHM solutions when we set $\chi = 0$.
In fact, in this case, the $\Omega$ field is set to a fixed
real value as can be seen from Eq. (\ref{omega}), since
$h_{\pm} = 0$ and the denominator is a constant for 
$A > \bar{A}_-$.  Thus, the topology of the space-time is 
${\cal M}_2 \times S^{d -2}$ where the radius of the transversal 
sphere is a constant.  This solution is drastically 
different from the
usual spherically symmetric 4-dimensional space-time, where
the $\Omega$ field is given by a non-constant function.

To summarize,  we recover the GHS (or GM) black holes for $0 < k < 1$
and the CHM black holes for $k = 1$.  As the black holes try to carry
the dilaton charge other than the value prescribed by $\bar{f}_0 = 0$,
naked singularities occur.  In this sense, the GHS  (or GM) black holes and
the CHM black holes are unique solutions where the essential singularities
are hidden inside the horizon.

\subsection{Solutions (37) }

The solutions given in Eq. (\ref{ops}) have 
an interesting space-time structure.
From Eq. (\ref{Aor}), we see that $P (A) / \bar{Q}$ should be chosen to be
positive semi-definite.   From Eq. (\ref{ops}), this implies $D^2 < 0$ case
gives physically acceptable solutions since we have $\mu < 0$ for the
theories in our consideration.  The sufficient 
but not necessary conditions to satisfy all the
restrictions are to require $D^2 < 0$ and $4 \bar{s}^2 - 2a \bar{Q} c < 0$.
Due to the latter condition, we see that $P{A} / \bar{Q}$ is 
positive definite
and, from Eq. (\ref{ops}), the range of $\Omega$ has a finite maximum 
value.  Thus, we have solutions
with a finite range of allowed radius of the transversal sphere.  The metric
looks particularly simple if we further impose $a f_0 + \chi s = 0$.
Then, we indeed  find 
\[ D^2 = \frac{a}{a + \chi^2 /2} ( 4 \bar{s}^2 - 2a \bar{Q} c ) < 0 \]
and
\[ \bar{f}_0^2 = - \frac{1}{2a} \frac{\chi^2 /2}{ a + \chi^2 /2} 
 ( 4 \bar{s}^2 - 2a \bar{Q} c ) > 0  \]
as long as $4 \bar{s}^2 - 2a \bar{Q} c < 0$.  For simplicity, we choose
$\bar{f}_0 > 0$.   Then in terms of geometric gauge, the metric can
be written as
\begin{equation}
ds^2 = \frac{4a \bar{f}_0^2 }{- \mu \chi^2}
\frac{dt^2}{r^{2a/(1-a)}}
- \frac{a +\chi^2 /2}{-\mu (1-a)^2}
\frac{dr^2}{ (r^* / r )^{2 (a + \chi^2 /2 )/a} -1 }
- r^2 d\Omega
\end{equation}
and the dilaton field is given by
\begin{equation}
f(r) = \frac{\chi}{1-a} \ln r - \frac{\chi }{2a} 
\ln ( \frac{4aQ \bar{f}_0^2}{-\mu \chi^2} )
+ \frac{a + \chi^2 /2}{a} f_1     .
\end{equation}
The constant $r^*$ is given by
\[ r^* = ( \frac{\chi^2 Q}{2a \bar{f}_0^2} 
 )^{\frac{a}{2a + \chi^2} }
\sqrt{\frac{4a Q \bar{f}_0^2 }{\chi^2 } e^{- \chi f_1 } } . \]
That this solution has a finite range of allowed value of the transversal
sphere is clear from the above expression for the metric.  
 
\section{Discussions}

We obtained general static solutions 
on a local space-time region
that continuously include the asymptotically
flat GHS (or GM) black holes and the non-asymptotically flat CHM
black holes.  We then demonstrated an analog of the no-hair theorem
for each black hole solution.  In this class of solutions, we mention
that $k > 1 $ solutions exactly reproduce the black hole solution with
$1/k < 1$.  This statement can be straightforwardly verified by setting
the range of $A$ to be $( \bar{A}_{\infty}^{I}  , - \bar{A}_+ ] $ and
repeating the calculations in section IIIA.   This shows a kind of duality
in our solutions and, incidently, the CHM solutions correspond to
the self-dual solution under this dual transformation.  It will be 
interesting to see whether there is some underlying reason for this
behavior.

We find, additionally, there are whole other class of
space-time with unusual geometries such as Eq. (\ref{ops}).  Even in 
Eq. (\ref{s1}), if we set $c_1 \le 0$, there are solutions with finite range
of $\Omega$ and asymptotically flat solutions with naked singularities.
Although all of these new solutions have unphysical feature, it is
still interesting to observe their unusual space-time geometries.
It remains to be seen whether we can construct non-trivial global
space-time structure that makes physical sense.

Our analysis in this paper is valid for $a >0$ as can be seen from
our field redefinitions.  This technical point excludes the case of
the (2+1)-dimensional gravity theories from our consideration.  The
(2+1)-dimensional gravity theories with a negative cosmological 
constant contain
black holes and have many interesting features \cite{btz}.
The treatment for this case in presence of the dilaton field
is in progress.

\acknowledgements{Dahl Park would like to thank Prof. Jae Kwan Kim for
his valuable guidance throughout this work.  We thank Prof. Dae Sung
Hwang for careful reading of this paper and other useful discussions.}

\end{document}